\documentclass[a4paper,11pt]{article}
\pdfoutput=1 % if your are submitting a pdflatex (i.e. if you have
\usepackage{jinstpub} % for details on the use of the package, please
\usepackage{eurosym}        

\title{Liquid xenon in nuclear medicine: state-of-the-art and the PETALO approach}

\author[a,b]{P. Ferrario}
\affiliation[a]{Donostia International Physics Center (DIPC),\\
Paseo Manuel de Lardizabal 4, 20018 Donostia-San Sebasti\'an, Spain}
\affiliation[b]{Instituto de F\'isica Corpuscular (IFIC), CSIC \& Universitat de Val\`encia,\\
Calle Catedr\'atico Jos\'e Beltr\'an, 2, 46980 Paterna, Valencia, Spain}
\emailAdd{paola.ferrario@ific.uv.es}

\abstract{Liquid xenon has several attractive features, which make it suitable for applications to nuclear medicine, such as high scintillation yield and fast scintillation decay time, better than currently used crystals. Since the '90s, several attempts has been made to build Positron Emission Tomography scanners based on liquid xenon, which can be divided into two different approaches: on one hand, the detection of the ionization charge in TPCs, and, on the other one, the detection of scintillation light with photomultipliers.

PETALO (Positron Emission Tof Apparatus with Liquid xenOn) is a novel concept, which combines liquid xenon scintillating cells and silicon photomultipliers for the readout. A first Monte Carlo investigation has pointed out that this technology would provide an excellent intrinsic time resolution, which makes it  possible to measure the Time-Of-Flight with high efficiency. Also, the transparency of liquid xenon to UV and blue wavelengths opens the possibility of exploiting both scintillation and Cherenkov light for a high-sensitivity TOF-PET.
}


\proceeding{LIDINE 2017: Light Detection In Noble Elements\\
 22-24 September 2017 \\
  SLAC National Accelerator Laboratory}

\begin{document}
\maketitle
\flushbottom

\section{Positron Emission Tomography}
\label{sec:pet}

\begin{figure}[htbp]
\centering % \begin{center}/\end{center} takes some additional vertical space
\includegraphics[width=\textwidth]{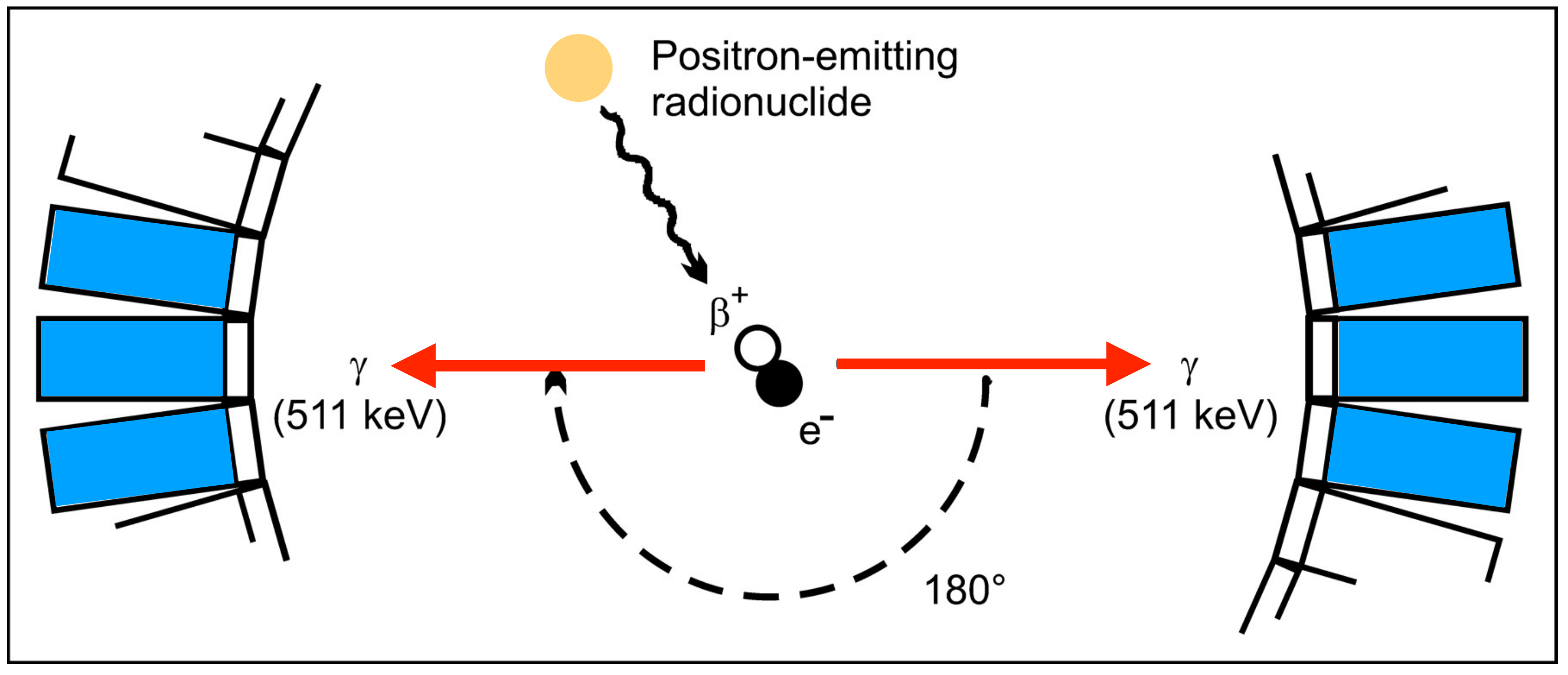}
\caption{\label{fig:lor} Scheme of the detection in coincidence of two back-to-back gammas in a PET scanner.}
\end{figure}

Positron Emission Tomography (PET) is a diagnostic imaging technique which scans the metabolic activity of the body. Measuring glucose metabolism, blood flow or oxygen use, it allows one to identify several diseases such as cancer, cardiovascular problems or brain disorders. This technique consists of injecting the patient with a molecule previously labeled with a radioactive isotope (radiotracer) that decays emitting a positron. The positron annihilates with an electron of the neighbouring atoms, producing two 511 keV photons with momenta on the same line (line of response, or LOR), but in opposite directions. A system of detectors surrounding the patient detects the coincidence of the two gammas, identifying the LOR (see figure~\ref{fig:lor}). Crossing many LORs allows one to reconstruct the image of the area where the radiotracer concentrates.

The performance of a PET scanner is defined by the physical properties of the detector and the characteristics of the readout. Its sensitivity is defined as the number of coincidences detected by the device per unit time per unit of activity of the source \cite{sensi}: 
\begin{equation}
\label{eq:sensi}
S=\frac{A\cdot \varepsilon^2 \cdot e^{-\mu t} \cdot \xi}{4\pi r^2}(\mathrm{cps}/\mu \mathrm{Ci})\,.
\end{equation}
The PET sensitivity depends on the detector area $A$ seen by the point source to be imaged, and on the detector efficiency $\varepsilon$ (which in turn depends on the detector scintillating time and the response of the sensors, multiplied by the fraction of events that are relevant for detection, such as photoelectric interactions). To increase the efficiency, one needs a photon yield per keV as high as possible and minimal losses of optical photons to record large signals, and an energy resolution as good as possible to reduce Compton coincidences inside the patient. The detector efficiency enters with a power of two, due to the need to form a coincidence with two detectors. The attenuation length to 511 keV gammas ($\mu$) sets the scale of the length (across the photon line of flight) that the detector needs to have in order to stop most of the incoming radiation; thus, materials with high density and high Z are preferred, or, otherwise, capable of compensating with larger thickness $t$ a longer attenuation length. Finally, $\xi = 3.7 \times 10^4$ is a numerical conversion factor and $r$ is the radius of the detector ring.

\section{The need for Time-Of-Flight}
\label{sec:tof}

\begin{figure}[htbp]
\centering % \begin{center}/\end{center} takes some additional vertical space
\includegraphics[width=0.7\textwidth]{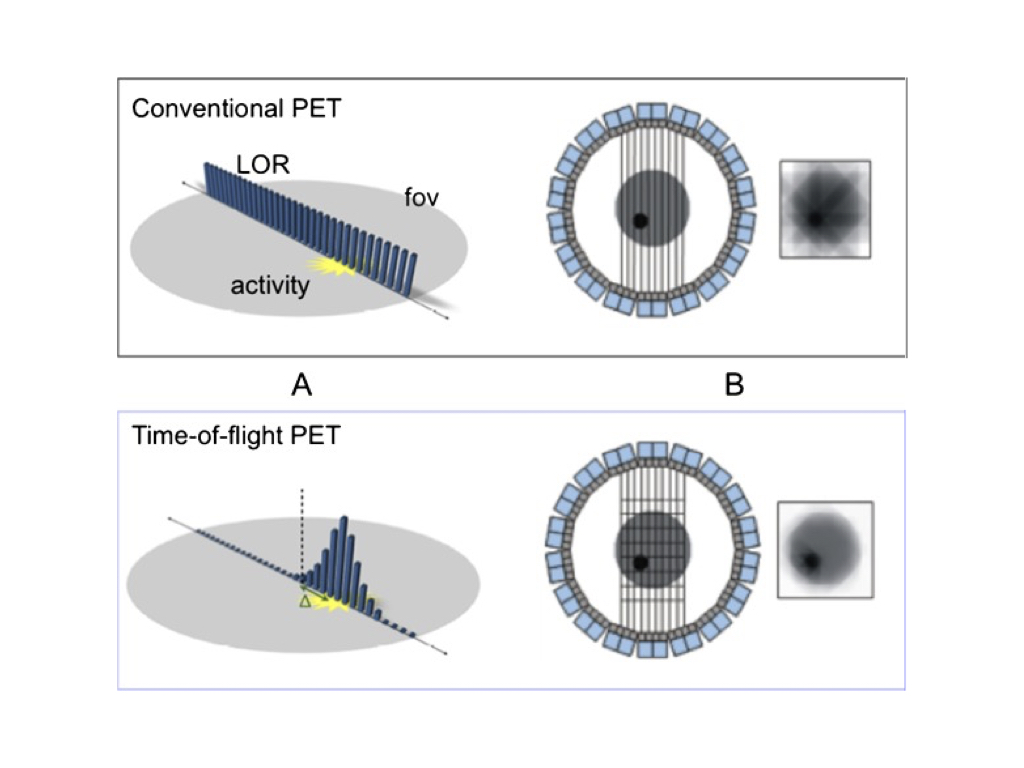}
\caption{\label{fig:tof} Illustration of the advantage of having time-of-flight measurement in a scanner. Adapted from ref.~\cite{beyer}.}
\end{figure}

The measurement of the time difference between the arrival of the two photons (time--of--flight, or TOF) makes possible to constrain to a larger extent the part of the LOR where the annihilation could have happened, thus increases the sensitivity of the scanner. The annihilation point is constrained to a narrower region on each LOR, therefore, the background in the image (meaning the voxels on the LOR which are attributed, erroneously, a probability of being the source of gammas) is reduced and the image contrast increases, as shown in figure~\ref{fig:tof}. An important consequence is that a smaller number of coincidences is required for the same image quality, thus the patient can receive a smaller dose of radiotracer. Moreover, the reconstruction time of the image decreases and this is an advantage for dynamic studies.

The gain due to TOF measurements is estimated to be proportional to $D/\Delta x$ \cite{tof}, where $D$ is the diameter of the object to be scanned and $\Delta x$ is the spatial resolution connected with TOF, i.e., $\Delta x = c \Delta t/2$, where $c$ is the speed of light and $\Delta t$ is the time resolution \cite{gain}. Since $c \sim$ 30 cm/ns, if one is able to measure $\Delta t$ to a resolution of 600 ps (corresponding to the typical time resolution achieved by current commercial systems), the resulting precision in the
determination of $\Delta x$ is around 9 cm, making TOF gain sizeable only for large object scans. However, a $\Delta t$ resolution of 25 ps would yield a resolution of better than 4 mm, and therefore TOF could be used to determine directly the emission point. 

%With TOF measurements, the noise pixels in the LOR are reduced to a narrower region, thus the signal to noise ratio is improved, as shown in figure~\ref{fig:tof}. An important consequence of this is that a smaller number of coincidences is required for the same quality of reconstruction, thus the patient can receive a smaller dose of radiotracer.
%

The resolution of TOF (known as coincidence resolving time, or CRT) is determined fundamentally by the time resolution of the readout and the scintillation decay time, which must be as short a possible, to maximize the number of events acquired per unit time and to minimize the window used to correlate events in different detectors. The CRT depends also on the transverse and longitudinal spatial resolution of the detector: a precise determination of the interaction point of gammas allows one to reduce the parallax error and the fluctuations in the measurement of the arrival time of the photons. Finally the CRT is also affected by the fluctuation of the speed of optical photons in the scintillation medium, which implies a fluctuation in the estimation of the arrival time. This fluctuation is due to the spread of the energy spectrum of the scintillation light of the medium.
%The time of flight difference between the two photons can be expressed as \cite{Gomez-Cadenas:2016mkq}:
%
%\begin{equation}
%\label{eq:CRT}
%\Delta t  \equiv 2\frac{\Delta x}{c}  =  t_1 - t_2 - \frac{\Delta d_g}{c} - \frac{\Delta d_p}{v_p}\,,
%\end{equation}
%%
%where $d_g$ is the distance from the centre of the geometrical system to the interaction vertex of each 511 keV gamma, $\Delta x$ is the displacement of the gamma emission vertex from the centre of the geometrical system and $d_p$ is the distance from the interaction vertex to the detection vertex (i.e., the position of the sensor that detects the first scintillation photon)

\section{Liquid xenon in PET imaging}
\label{sec:state-of-the-art}

Xenon is a noble gas, which responds to the interaction of ionizing radiation producing about 60 photons per keV of deposited energy \cite{Chepel:2012sj}, in the absence of electric field. When a 511 keV gamma interacts in liquid xenon (LXe) it produces a secondary electron (by either photoelectric effect, which happens in $\sim$20\% of the events, or Compton interaction) which in turn propagates for a short distance in the liquid, ionizing and exciting the medium.  The emitted scintillation photons have wavelengths in the ultraviolet range with an average of 178 nm.  In its liquid phase xenon has a reasonably high density (2.98 g/cm$^3$ \cite{DensityLXe}), which gives an attenuation length of 3.7 cm for 511-keV gammas \cite{AttLengthLXe}  and an acceptable Rayleigh scattering length (29 cm \cite{RayleighLXe}), which makes it suitable for PET applications. The main attractive features of liquid xenon (LXe) are a high scintillation yield  ($\sim$ 30\,000 photons per 511 keV gamma), and a fast scintillation decay time  (2.2 ns in its fastest mode) .

%and the transparency to its own Cherenkov light. Cherenkov radiation is produced promptly (few picoseconds compared to the nanoseconds of scintillation), thus is the best candidate for TOF measurements.  However, the yield is lower than that of scintillation, thus the efficiency of detecting two photons in coincidence is very low. The distribution of Cherenkov light in terms of wavelengths follows a $1/\lambda^2$ trend, thus most of Cherenkov light is produced in the UV range. Typical crystals such as LSO have drop in transparency for wavelengths lower than 400 nm \cite{nLYSO}, thus most of Cherenkov light is not detected. On the contrary, xenon has the advantage of being transparent to UV and blue light, which increases light collection and, therefore, the rate of detection of coincidences.

Liquid xenon is a continuous medium with uniform response. Therefore, the design of a compact system is much simpler than in the case of solid detectors of fixed shape. It is also possible to provide a 3D measurement of the interaction point, and, thus, a high resolution measurement of the depth of interaction (DOI). Furthermore, in LXe it is possible in principle to identify Compton events depositing all their energy in the detector as separate-site interaction, due to its relatively large interaction length.  
%Once an event in the region of interest (around 511 keV of total deposited energy) is identified, the pattern of recorded light on SiPMs can in principle be inspected to find one or more depositions, using, for instance, neural networks \cite{tfm}. 
This increases the sensitivity of the system, since those events can be used for image reconstruction in addition to the photoelectric ones. Furthermore, the temperature at which xenon can be liquefied at a pressure very close to the atmospheric (the triple point of xenon is reached at a temperature of 161.35 K and a pressure of 0.816 bar \cite{TempLXe})  is high enough as to be reached using a basic cryostat. %Also, at this temperature SiPMs can be operated normally, and their dark count rate is essentially negligible (see, for instance, figure 17, in reference \cite{dcr}). 
Finally, LXe is cheaper than current crystals (around 3 \euro/cc to be compared with 40--50 \euro/cc for LSO). 

The possibility of building a PET based on the excellent properties of LXe as scintillator was first suggested by Lavoie in 1976 \cite{lavoie}. In 1993, Chepel proposed to read both scintillation and ionization charge in a LXe Time Projection Chamber (TPC)\cite{chepelFirst} and an extensive work has been done since then by him and co-workers \cite{chepelLXe, chepelRes, chepelEnergyRes}. In the late '90s, Doke and Masuda proposed to use a liquid xenon scintillating calorimeter read out by photomultipliers surrounding completely the whole volume \cite{DokeMasuda}. This idea, applied to PET, was studied by the Waseda group \cite{Doke1,Nishikido2,Nishikido1} with a prototype based in LXe cells instrumented with VUV-sensitive PMTs which covered 5 of the 6 sides of the cell.  In 2007, the Xemis group at Subatech started building a LXe Compton telescope, which uses a $\beta$+ decay isotope that also emits a high energy gamma, such as $^{\ensuremath{44}}$Sc \cite{xemis1, GallegoManzano:2015hkg}. While the coincidences of the two annihilation photons are detected by a a conventional PET ring, the high energy gamma converts through Compton interaction in a LXe TPC. The Compton kinematics allows one to reconstruct the direction of the gamma, therefore to constrain the region of the LOR where the decay occurs.

%None of these studies has originated a commercial PET scanner based on LXe.

These studies didn't exploit all the potential of LXe applied to PET. On one hand, although reading both light and charge gives a better energy and spatial resolution, a TPC is slow and introduces dead time due to charge drift, which may decrease the sensitivity and increases the complexity and cost of the detector, thus jeopardizing the large scale implementation of the technology. On the other hand, the large size of PMTs results in less precise geometric corrections and a worse spatial resolution. 
%Spatial resolution is also affected by leaving the entry face of the cell uninstrumented, since this is the closer face to the interaction point of the 511-keV gamma, thus it is the one that gives the best resolution in its reconstruction. The choice of leaving it without instrumentation was forced by the large mass of PMTs, which would have blocked most gammas.

\section{The PETALO approach}
\label{sec:petalo}

A new approach has been proposed recently \cite{Gomez-Cadenas:2016mkq, Gomez-Cadenas:2017bfq}, which is based on liquid xenon scintillating cells. The key idea of PETALO (a Positron Emission TOF Apparatus based on Liquid xenOn) is to capture with high efficiency, minimal geometrical distortions and uniform response most of the light produced by the scintillation of xenon, in order to achieve good energy and space resolution and perform TOF measurements. To this end, the sensors chosen for the light readout are SiPMs, which provide large area, high gain and very low noise. At the liquefaction temperature of xenon, SiPMs can be operated without problems, and their dark count rate is essentially negligible. In its simplest version, each cell has the entry and the exit faces instrumented, while the others are covered with a highly reflective material such as PTFE, which features close to 100\% reflective efficiency at UV wavelenghts \cite{reflPTFE}. The shape and the dimensions of the box itself can be adapted to the specific application, as well as the density of SiPMs on each face. A first Monte Carlo study has shown that a very good CRT of down to 70 ps FWHM can be reached using VUV sensitive SiPMs with a photodetection efficiency of 20$\%$ \cite{Gomez-Cadenas:2016mkq}.

This approach overcomes the weak points of previous attempts to use LXe in PET technology, since no high voltages and dead times due to drift are needed. Moreover, it exploits new generation sensors, which are smaller, therefore can be used to instrument also the entry face of the cell, thus providing a better spatial resolution in the reconstruction of the gamma interaction.

A very interesting aspect of the PETALO concept is the possibility of pursuing a Cherenkov-based PET. Using Cherenkov light for TOF is very attractive, due to the promptness of these photons, but has the drawback of a low yield and a high absorption rate in conventional crystal detectors. The distribution of Cherenkov light in terms of wavelengths follows a $1/\lambda^2$ trend, thus most of Cherenkov light is produced in the UV range. Typical crystals such as LSO have a drop in transparency for wavelengths lower than 400 nm \cite{nLYSO}, thus most of Cherenkov light is not detected. %Recent studies with PbF$_2$ crystals instrumented with microchannel plates have shown a rather low efficiency, due to the small number of photons detected in coincidence \cite{KorparSiPMs}.  
On the contrary, xenon has the advantage of being transparent to UV and blue light, which increases light collection and, therefore, the rate of detection of coincidences. A first study has been published \cite{Gomez-Cadenas:2017bfq} which shows that a CRT of 30 -- 50 ps can be achieved, using very fast photosensors, sensitive to wavelengths down to $\sim$ 300 nm, such as microchannel plates.

\acknowledgments

Support from GVA under grant PROMETEO/2016/120 is acknowledged.

%\paragraph{Note added.} This is also a good position for notes added
%after the paper has been written.


\bibliography{biblio}

%\begin{thebibliography}{99}

%\bibitem{a}
%Author, \emph{Title}, \emph{J. Abbrev.} {\bf vol} (year) pg.
%
%\bibitem{b}
%Author, \emph{Title},
%arxiv:1234.5678.
%
%\bibitem{c}
%Author, \emph{Title},
%Publisher (year).

% Please avoid comments such as "For a review'', "For some examples",
% "and references therein" or move them in the text. In general,
% please leave only references in the bibliography and move all
% accessory text in footnotes.

% Also, please have only one work for each \bibitem.

%\end{thebibliography}
\end{document}